\documentclass[letterpaper, 10 pt, conference]{ieeeconf}  

\IEEEoverridecommandlockouts                              

\usepackage[utf8]{inputenc}
\usepackage[T1]{fontenc}
\usepackage{algorithm,algorithmic}
\usepackage{tabularx,booktabs}
\usepackage{graphics}
\usepackage{float} 
\usepackage{epsfig}
\usepackage{mathtools}
\usepackage{amsfonts}       
\usepackage{mathtools}
\usepackage{multirow}
\usepackage{graphicx}

\title{\LARGE \bf
Benchmarking Domain Generalization on EEG-based Emotion Recognition
}

\author{Yan Li$^{1,\dagger}$, Hao Chen$^{2,3,\dagger}$, Jake Zhao (Junbo)$^{1}$, Haolan Zhang$^{1,4}$, and Jinpeng Li$^{2,3, *}$,~\IEEEmembership{Member,~IEEE,} 
\thanks{This work was supported in part by National Natural Science Foundation of China (62106248), Zhejiang Provincial Natural Science Foundation of China (LQ20F030013), and Ningbo Public Service Technology Foundation, China (202002N3181).}
\thanks{$^{\dagger}$ \textit{equal contribution}}
\thanks{$^{1}$ School of Computer Science, Zhejiang University, Hangzhou, Zhejiang, China.}
\thanks{$^{2}$ HwaMei Hospital, University of Chinese Academy of Sciences, No. 41 Northwest Street, Haishu District, Ningbo, Zhejiang, 315010, China.}
\thanks{$^{3}$ Ningbo Institute of Life and Health Industry, University of Chinese Academy of Sciences, Ningbo, Zhejiang, China.}
\thanks{$^{4}$ NIT, Zhejiang University, Ningbo, Zhejiang, China.}
\thanks{$^{*}$ Corresponding author: Jinpeng Li (\textit{E-mail: lijinpeng@ucas.ac.cn})}
}

\begin{document}

\maketitle
\thispagestyle{empty}
\pagestyle{empty}

\begin{abstract}
Electroencephalography (EEG) based emotion recognition has demonstrated tremendous improvement in recent years. Specifically, numerous domain adaptation (DA) algorithms have been exploited in the past five years to enhance the generalization of emotion recognition models across subjects. The DA methods assume that calibration data (although unlabeled) exists in the target domain (new user). However, this assumption conflicts with the application scenario that the model should be deployed without the time-consuming calibration experiments. We argue that domain generalization (DG) is more reasonable than DA in these applications. DG learns how to generalize to unseen target domains by leveraging knowledge from multiple source domains, which provides a new possibility to train general models. In this paper, we for the first time benchmark state-of-the-art DG algorithms on EEG-based emotion recognition. Since convolutional neural network (CNN), deep brief network (DBN) and multilayer perceptron (MLP) have been proved to be effective emotion recognition models, we use these three models as solid baselines. Experimental results show that DG achieves an accuracy of up to 79.41\% on the SEED dataset for recognizing three emotions, indicting the potential of DG in zero-training emotion recognition when multiple sources are available.

\end{abstract}

\section{INTRODUCTION}

Emotion recognition is of great importance for humans in various aspects of daily activities. In human-computer interaction, emotion plays an essential role in fatigue detection and healthcare, and existing studies have confirmed the association between various diseases and emotions~\cite{valstar2016avec}. Human emotions can be detected via facial expression, speech and eye blinking~\cite{sariyanidi2014automatic}, etc. However, these methods are always susceptible to subjective influences of the participants, which will affect the accuracy and reliability of the models. Comparatively, recognizing emotions through physiological signals is more objective and reliable. As a bridge between the brain and the computer, brain-computer interfaces (BCIs) allow users to acquire brain signals directly. Invasive BCIs, for example, are prohibitively expensive and need surgery to achieve a better accuracy. On the other hand, non-invasive BCIs using electroencephalography (EEG) are much safer and thus have been commonly employed to collect brain signals~\cite{cincotti2008non}. Typically, feature extraction and classification are employed on the preprocessed EEG data.

Nevertheless, the EEG signals acquired from the same subject at the same session can be very biased, and training a general model remains a challenge in EEG-based emotion recognition. In recent years, to tackle this issue, numerous works have applied transfer learning, especially domain adaptation (DA) to transfer knowledge from the labeled source domain (existing subjects) to the unlabeled target domain (new data). For example, Zheng~\textit{et al.} applied transfer component analysis (TCA) to help improve the accuracy in the cross-subject transfer~\cite{zheng2016personalizing}. Li~\textit{et al.} proposed the multi-source style transfer mapping (MS-STM) for the cross-subject multi-source scenario to reduce the marginal distribution differences~\cite{li2019multisource}.
With the development of deep learning, Li~\textit{et al.} adopted deep adaptation network for cross-subject emotion recognition~\cite{li2018cross}. Zheng~\textit{et al.} extended the SEED dataset to SEED-IV with more emotion categories and additional modality of emotion data. They also presented EmotionMeter, which fuses two modalities of EEG data~\cite{zheng2018emotionmeter}. 
Zhao~\textit{et al.} proposed a plug-and-play domain adaptation method for shortening the calibration time while maintaining the accuracy~\cite{zhao2021plug}. Chen~\textit{et al.} focused on the multi-source scenarios in EEG-based emotion recognition and presented MS-MDA~\cite{chen2021meernet, chen2021ms} to take both domain-invariant and domain-specific EEG features into consideration.

Most of the aforementioned works enhanced the performance of the model on the target domain by decreasing the domain shift between the source and target domain. However, the assumption that unlabeled target data exists conflicts with practical application, which does not require additional data acquisition in the target domain. Besides, the cost of training a model using DA for every single target domain is high. Domain generalization (DG), on the other hand, deals with a more challenging setting that several different but related domain(s) are given, and the goal is to learn a zero-training model that can generalize to the unseen test domain~\cite{2021Generalizing}. DG helps extract domain-invariant features by exploiting domain differences across multiple source subjects without acquiring any extra target data. However, there are limited works using DG in EEG-based emotion recognition~\cite{ma2019reducing}. Benchmarking DG methods on EEG-based emotion recognition is necessary to provide insights and references for the community.

In this paper, we benchmark state-of-the-art DG algorithms on EEG-based emotion recognition, which helps alleviate the differences of multiple domain distributions to achieve better generalization on unseen subjects. For each DG method, we use strong baselines for training, including convolutional neural network (CNN), deep brief network (DBN) and multilayer perceptron (MLP), which have been demonstrated effective in emotion recognition. We evaluate the performance of these approaches by recognizing three emotions in a zero-training scenario.

\section{METHOD}
\begin{figure}
    \centering
    \includegraphics[width=\linewidth]{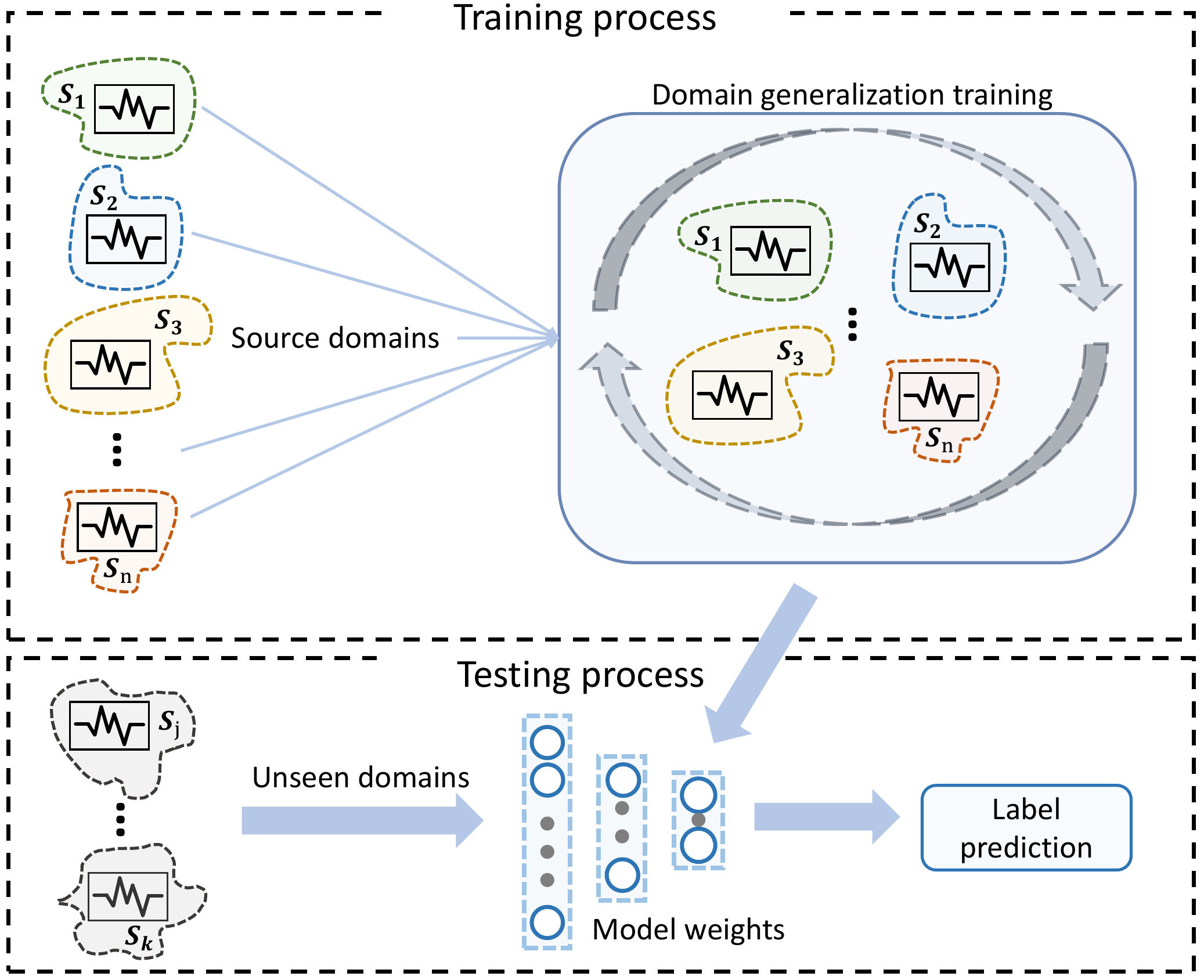}
    \caption{The overall {\bf framework} of domain generalization in EEG emotion recognition. Data from each subject is seen as a domain. In the training process, all the source domains are used to train a feature extractor by domain generalization methods. In the testing process, the model trained is used to extract the features from unseen domains and then predict the emotion labels. This figure is best viewed in colors.}
    \label{fig:framework}
\end{figure}

The overall framework of the application of domain generalization in EEG-based emotion recognition is presented in Fig. \ref{fig:framework}. Data from each subject can be seen as a domain, and the unseen subject data domains are taken as the test sets for the inference stage.

Based on the aspects that are focused on during training, the selected DG methods can be categorized into three types.

\subsection{Data Manipulation}
We select two data manipulation approaches named Mixup~\cite{zhang2017mixup} and group DRO~\cite{sagawa2019distributionally}. Mixup extends the training distribution by incorporating the prior knowledge that linear interpolations of feature vectors should lead to linear interpolations of the associated targets. The group DRO leverages prior knowledge of spurious correlations to define groups over the training data and define the uncertainly set in terms of these groups.

\textbf{Mixup:} It generates virtual feature-target vectors from real feature-target vectors.
The whole Mixup method can be described as
\begin{equation}
\begin{aligned}
  \mu(\tilde{x},\tilde{y}|x_i,y_i) = \frac{1}{n}\sum_{j}^{n}\mathbb{E}[\delta(\lambda x_i + (1-\lambda)x_j, \lambda y_i + (1-\lambda)y_j)],
\end{aligned}
\end{equation}
where $\mathbb{E}$ represents the empirical risk minimization (ERM)~\cite{vapnik1999nature}. $(x_i,y_i)$ and $(x_j,y_j)$ are two feature-target vectors drawn at random from the training data, $\lambda \in [0,1]$. We minimize the average of the loss function over the data distribution to train the model:
\begin{equation}
Min(R(f)) = Min(\int l(f(x),y)dP(x,y)),
\end{equation}
where $l$ denotes the loss function and $P$ denotes the data distribution.

\textbf{Group DRO:} It needs to define $m$ groups $P_g$ indexed by $G ={1,2,...,M}$ and the training distribution $P$ is a mixture of the $m$ groups. The uncertainty set $Q$ is defined by any mixtures of these groups:
\begin{equation}
Q = {\sum_{m}^{g=1}q_gP_g, \quad q\in \triangle_m},
\end{equation}
where $\triangle_m$ is the (m-1)-dimensional probability simplex. The group DRO method yields the model by minimizing the empirical worst-group risk $\hat{R}(\theta)$:
\begin{equation}
\begin{aligned}
\hat{\theta}_{DRO} = min{\hat{R}(\theta) = max \mathbb{E}_{(x,y)\in P_g}[l(\theta;(x,y))]}.
\end{aligned}
\end{equation}
$\mathbb{E}$ represents ERM method and $l$ represents the loss function.

\subsection{Representation Learning}
Deep domain confusion (DDC)~\cite{tzeng2014deep}, deep adversarial neural network (DANN)~\cite{ganin2015unsupervised}, and deep CORAL~\cite{sun2016deep} from selected methods belong to the representation learning. DANN embeds the domain adaptation into the process of learning representation, so that the final classification decisions are made based on features that are both discriminative and invariant to the change of domains. DDC uses an adaption layer along with a domain confusion loss based on maximum mean discrepancy (MMD)~\cite{borgwardt2006integrating} to automatically learn a representation jointly trained to optimize for classification and domain invariance. The deep CORAL constructs a differentiable loss function that minimizes the difference between source and target correlations.

The DANN method try to make the two feature distributions as similar as possible and discriminate the two feature distribution at the same time, which can be represented as
\begin{equation}
\begin{aligned}
E(\theta_f,\theta_y,\theta_d) = 
\sum_{i=1,..,N}L_{y}^i(\theta_f,\theta_y)-\lambda\sum_{i=1,..,N}L_{d}^i(\theta_f,\theta_d),
\end{aligned}
\end{equation}
where $\theta_f$ denotes the parameters of the feature mapping that maximize the loss of the domain classifier; $\theta_d$ is the parameter of the domain classifier that minimize the loss of the domain classifier; $L_y$ is the loss for label prediction and $L_d$ is the loss for domain classification. $L_{y}^i$ and $L_{d}^i$ denote the corresponding loss function evaluated at the $i-th$ training example.

The DDC method trains the classifier on the labeled source data through minimizing the distance between the source and target distributions, and a standard distribution distance metric MMD is used, which is defined as
\begin{equation}
MMD(X_S,X_T) = 
\|\frac{1}{|X_S|}\sum_{x_s\in X_S}\phi(x_s)  - \frac{1}{|X_T}\sum_{x_t\in X_T}\phi(x_t)\|,
\end{equation}
where $X_S$ denotes the source data and $X_T$ denotes the target data, and $\phi$ represents a kernel function. The final loss function can be written as
\begin{equation}
l = l_C(X_L,y) + \lambda MMD^2(X_S,X_T),
\end{equation}
where $l_c(X_L,y)$ denotes classification loss on the available labeled data $X_L$ and the ground truth label$y$, $\lambda$ is a hyper-parameter.

The deep CORAL defines a CORAL loss:
\begin{equation}
l_{CORAL} = \frac{1}{4d^2}\|C_S - C_T\|_{F}^{2},
\end{equation}
where $\|\cdot\|_{F}^{2}$ denotes the squared matrix frobenius norm, and $C_S, C_T$ are generated by the labeled source-domain data and  unlabeled target data. 
And in order to learn features that work well on the target domain, both the classification and CORAL loss need to be trained at the same time.

\subsection{Learning Strategy}
We adopt the RSC~\cite{huang2020self} to train the models. The RSC method discards the representation associated with the higher gradients at each epoch and forces the model to predict with remaining information during the training process.

RSC first calculates the gradient of upper layers with respect to the representation
\begin{equation}
g_z = \partial(h(z,\hat{\theta}_{t}^{top})\odot y)/\partial z,
\end{equation}
where $\odot$ denotes an element-wise product; $z$ is the feature representation; $h(z,\hat{\theta}_{t}^{top})$ denotes the task component of the model with input $z$ and parameter $\hat{\theta}_{t}^{top}$; $t$ denotes the $t-th$ iteration.
The gradient can be written as below:
\begin{equation}
 \tilde{g}_\theta = \partial l(\sigma(h(\tilde{z};\hat{\theta}_{t}^{top})),y)/\partial \hat{\theta}_{t}.
\end{equation}

\section{EXPERIMENTS}
We perform emotion recognition tasks on the SEED dataset~\cite{duan2013differential, zheng2015investigating} with all the considered DG methods. We also experiment with different baselines to extract features, including well-tuned CNN, DBN and MLP. The Institution’s Ethical Review Board approved all experimental procedures involving human subjects.

\subsection{Settings}
{\bf Datasets:} The SEED dataset contains EEG signals from 15 healthy subjects (7 males and 8 females) with three emotion categories (negative, neutral, and positive). Each subject performed the signal acquisition three times with an interval of one week, and the trial number of each session is 15. The raw data are recorded with an ESI NeuroScan system with 62 channels, and then down-sampled to 200 Hz sampling rate. After that, a band-pass filter between 0-75 Hz was applied to maintain informative bands. There are several kinds of features extracted from raw data in SEED, among them, differential entropy (DE)~\cite{duan2013differential} has been proven to be robust and accurate.
It should be noticed that for each emotion category, samples should resized into the same shape to facilitate the input of the model. Here we set the input shape to (62, 250, 5) and add 0 to samples that are smaller than this size.

{\bf Implementation Details. } 
The Adam optimizer is used and the initial learning rate is set to $1e-2$. The batch size is set to 32, the model is trained with 50 Epochs with $5e-4$ weight decay. For the lambda of Mixup, we simply set to 0.2. The drop factor of RSC is set to 1/3. All the other parameters that are required for DG methods are set to 1 by default.
%

We select one subject as the unseen target domain and the remaining domains are divided into training domains and validation domains according to a ratio of 4:1. During the training process, the model with the highest accuracy on the validation domains is saved, and then we apply it to the target domain to compute the accuracy on the target domain. In order to avoid the contingency of the experiment, each subject is selected to be the target domain in turn. The mean value and the variance of the total 15 results are reported.

To show the effectiveness of our model, we also compare several different baselines: \textbf{ResNet}~\cite{he2016deep} (ResNet18, ResNet34, ResNet50). \textbf{MLP} (MLP model with 2, 3, and 4 fully connected layers here for experiment). \textbf{DBN} (We use a DBN model with two RBM layers, and the hidden units of the two RBM layers are 23*23*5 and 18*18*2).

\begin{table}[!t]
\normalsize
\centering
    \caption{The comparison of accuracy on SEED dataset shown in mean/std. The average accuracy numbers of each baseline and each DG method are given. The best results are in bold.}
    \label{tab:results}
    \resizebox{\linewidth}{!}{%
    \begin{tabular}{cclllllll}
        \toprule
        \textbf{baseline}                              & \multicolumn{1}{l}{} & ERM & DANN & RSC & Mixup & MMD & CORAL & average \\ \midrule
        \multicolumn{1}{c}{\multirow{2}{*}{ResNet-18}} & mean & 0.6429 & 0.5733 & 0.6178 & 0.7378 & 0.6859 & 0.6889 & 0.6578 \\
        \multicolumn{1}{c}{}                           & std & 0.1153 & 0.1194 & 0.1215 & 0.1005 & 0.0839 & 0.1375 & 0.1130 \\ \midrule
        \multirow{2}{*}{ResNet-34}                     & mean & 0.4770 & 0.5111 & 0.5318 & 0.7067 & 0.6607 & 0.6415 & 0.5881 \\
                                                       & std & 0.0979 & 0.1460 & 0.1300 & 0.0802 & 0.1180 & 0.1065 & 0.1131 \\ \midrule
        \multirow{2}{*}{ResNet-50}                     & mean & 0.4207 & 0.4163 & 0.4163 & 0.4844 & 0.5867 & 0.5852 & 0.4849 \\
                                                       & std & 0.1030 & 0.0967 & 0.1068 & 0.1070 & 0.1251 & 0.1292 & 0.1113 \\ \midrule
        \multirow{2}{*}{MLP-2}                         & mean & 0.7718 & 0.7397 & 0.7674 & 0.7807 & 0.7674 & 0.7333 & 0.7600 \\
                                                       & std & 0.0859 & 0.1104 & 0.0852 & 0.0712 & 0.1234 & 0.1197 & 0.0993 \\ \midrule
        \multirow{2}{*}{MLP-3}                         & mean & 0.7703 & 0.7572 & 0.7585 & 0.7600 & 0.7304 & 0.7363 & 0.7521 \\
                                                       & std & 0.0915 & 0.0661 & 0.0983 & 0.0798 & 0.1025 & 0.1257 & 0.0940 \\ \midrule
        \multirow{2}{*}{MLP-4}                         & mean & 0.7837 & 0.7896 & \textbf{0.7941} & 0.7733 & 0.757 & 0.763 & \textbf{0.7768} \\
                                                       & std & 0.0981 & 0.1108 & 0.1047 & 0.1002 & 0.1017 & 0.1248 & 0.1067 \\ \midrule
        \multirow{2}{*}{DBN}                           & mean & 0.6563& 0.6800 & 0.6533 & 0.7393 & 0.7244 & 0.7244 & 0.6690 \\
                                                       & std & 0.0605 & 0.0722 & 0.0796 & 0.0597 & 0.0818 & 0.0818 & 0.0683 \\ \midrule
        \multirow{2}{*}{}                              & \multirow{2}{*}{average} & 0.6461 & 0.6382 & 0.6485 & \textbf{0.7117} & 0.7018 & 0.6961 & \\
                                                       & & 0.0932 & 0.1031 & 0.1037 & 0.0855 & 0.1052 & 0.1179 \\
        \bottomrule
    \end{tabular}
    }
\end{table}

\subsection{Results} \label{results}

\begin{figure}[!t]
    \centering
    \includegraphics[width=1\linewidth]{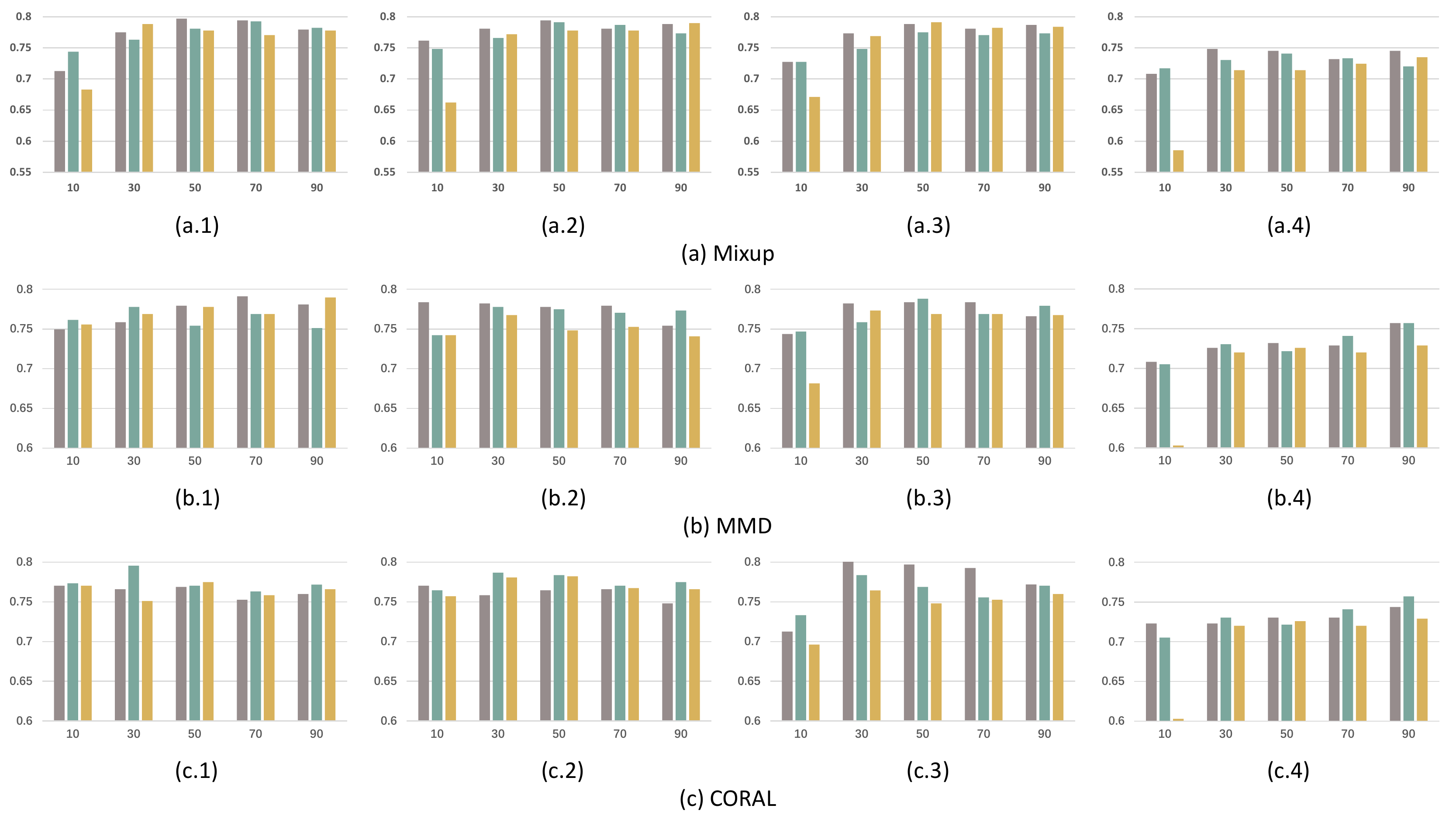}
    \caption{The influences of epoch and batch size on SEED dataset with three representative DG methods (Mixup, MMD, CORAL). Gray bar stands for batch size of 8, green bar stands for batch size of 16, and yellow bar stands for batch size of32. (*.1) is MLP-2 baseline, while (*.2) is MLP-3, (*.3) is MLP-4, and (*.4) is DBN.}
    \label{fig:additional}
    \vspace{3pt}
\end{figure}

Table \ref{tab:results} shows the results on SEED with different DG algorithms and baselines. For CNN, it can be seen that with the number of convolutional layers increases, the performance of ResNet series models decreases. As for the MLP models, there is no significant change in performance with increasing number of layers. Besides, all MLP models outperform ResNet series models, which indicates that CNNs may not be the best choice for EEG-based emotion recognition since the input is quite different from images. The results of DBN shows that DBN outperforms ResNet series but not as good as MLPs. Among these results, the highest accuracy is 0.7941 with RSC method using MLP-4 baseline, the best DG method is Mixup with an average on all baselines of 0.7117, and the best baseline is MLP-4 with an average on all DG methods of 0.7768. We also investigate the influences of epoch and batch size on three representative DG methods (Mixup, MMD, CORAL) with four outstanding models (MLP-2, MLP-3, MLP-4, DBN), the results are shown in Fig. \ref{fig:additional}.

In general, these results indicate that DG algorithms combined with suitable baselines can largely reduce the individual differences. The DG algorithms have broad application prospects in EEG-based tasks and provide a new direction for the broader application of EEG-based tasks.

The main insufficiency of this work is that we only consider deep learning methods that have been used more often in recent years and do not systematically test the DG algorithms on traditional classifiers such as support vector machines and linear discriminative analysis. When the data amount of source is small, traditional models may have better results. This is the next step of our work plan.

\section{CONCLUSIONS}
In this paper, we evaluate six representative domain generalization methods on SEED dataset with deep baselines, i.e., CNN, DBN and MLP. Based on experimental evaluations, we find that the well-tuned MLP can reach an accuracy of 79.41\% with RSC method when using no target data. Besides, all the results show that DG algorithms combined with specific baselines have the ability to achieve prominent effects on EEG-based emotion recognition for new users. The benchmarked DG method seems a promising routine towards zero-training emotion recognition models. It is suitable to the practical scenario of affective BCIs and serves as an inspiration and reference for subsequent works concerning transfer learning.

\bibliographystyle{ieeetr}
\bibliography{root}

\end{document}